
\noindent{{\bf Astrophysical Journal, 462, 489 (1996)\vskip-20pt}

\input cp-aa.tex



\overfullrule=0pt

\def\bfrm{\bf}

\def\hf{\hfill}
\newdimen\thicksize
\newdimen\thinsize
\thicksize=1.8pt
\thinsize=0.6pt
\def\th{\thinspace}

\def\qquad{\quad\quad}
\def\ngth{\negthinspace}

\def\hf{\hfill}

\def\frac#1#2{{\textstyle{ #1 \over #2}}}
\def\Apriori{{\it A priori}}
\def\apriori{{\it a priori}}

\def\at{{\rm\char'100}}
\def\eg{{{\it e.g.}\ }}
\def\etal{{\it et al.\ }}
\def\etc{{\it etc.\ }}

\def\cf{{\it cf.\ }}

\def\ie{{{\it i.e.}\ }}

\def\viz{{\it viz.\ }}
\def\vs{{\it vs.\ }}
\def\qv{{\it q.v.}}

\def\ni{\noindent}

\def\Mo{{$M_\odot $}}

\def\bF{{\bf F}}
\def\bG{{\bf G}}
\def\bH{{\bf H}}
\def\bX{{\bf X}}
\def\bY{{\bf Y}}
\def\bZ{{\bf Z}}
\def\approxgt{\raise4pt \hbox{$>$}\kern-9pt\lower1.5pt\hbox{$\sim$}}
\def\approxlt{\raise4pt \hbox{$<$}\kern-9pt\lower1.5pt\hbox{$\sim$}}

\def\Vec{\mathaccent"017E }
\def\vecr{{\Vec r}}
\def\vecu{{\Vec u}}

\MAINTITLE{Nonlinear Analysis of the Lightcurve of the Variable Star R~Scuti}

\AUTHOR{
J. R. Buchler\th@1\FOOTNOTE
{e--mail address: buchler\at phys.ufl.edu\hfill},
Z. Koll\'ath\th@1\FOOTNOTE
{on leave from Konkoly Observatory, Budapest,Hungary},
T. Serre,\th @1
\&
J. Mattei\th@2}

\INSTITUTE{
@1 Physics Department, University of Florida, Gainesville, FL 32611
@2 AAVSO, Cambridge, MA 02138
}

\DATE{30 september}

\ABSTRACT{It is first shown that the observational light curve data of R~Scuti,
a star of the RV~Tau type, is not multi-periodic, and that it cannot have been
generated by a linear stochastic (AR) process.  By default, the signal must be
a manifestation of deterministic chaos.  We use a novel nonlinear time-series
analysis, the global flow reconstruction technique, to probe the properties of
the irregular pulsation cycles.  We show in particular that the chaotic
dynamics of this star's complicated lightcurve is captured by a simple 4D
polynomial map or flow (4 first order ODEs).

Importantly also, the method allows us to {\sl quantify} an irregular signal
which has the potential benefit for extracting novel stellar constraints from
an irregular light-curve.  

Finally, from the low dimensionality 4 of the flow we can infer a simple
physical picture of the pulsations, and arguments are presented that the
pulsations of R Sct are the result of the nonlinear interaction of two
vibrational normal modes of the star.}


\maketitle

\input psfig

\def\rahmen#1{}

\titlea{\ \ INTRODUCTION}

It has been known for a long time that the metal-poor, Population II Cepheids,
labelled W~Virginis and RV~Tauri stars, often pulsate irregularly (\eg
Ludendorff 1928; Kukarkin 1975).  In fact, the pulsations of the W~Vir stars
which are of lower period and lower luminosity, are essentially regular
(periodic), but as the periods increase alternations appear in the oscillations
which however remain small for the W~Vir stars, \ie up to about 35 days (\eg
Arp 1955).  At still higher periods, \ie in the RV~Tau regime, the pulsations
become more irregular, with alternating very deep and shallow minima and with
large, long-term modulations in the amplitude.

The little theoretical work on these stars has been confined to their
evolution.  They are believed to be low mass, typically 0.6 -- 0.8\Mo\ which,
after helium exhaustion in the core, evolve to the red to the second giant
branch and then enter the instability strip (Gingold 1974).  Essentially no
theoretical attention had been paid to their pulsations because of the want of
a credible mechanism for the irregularity.  Some eight years ago the numerical
hydrodynamical {\sl modelling} of W~Vir stars of Buchler \& Kov\'acs (1987,
hereafter BK87; see also Kov\'acs \& Buchler 1988, hereafter KB88; reviewed in
Buchler 1990) showed that the irregularity in the pulsations is the
manifestation of {\sl low-dimensional chaos} and that it is no longer necessary
to invoke a {\it deus ex machina}, such as, for example, postulated irregular
convective phenomena.  (We recall that chaos occurred in the purely radiative
hydro-models.)  Unfortunately, at the time, it was not possible to support
these theoretical conclusions as to the chaotic nature of the pulsations with
an analysis of {\sl~observational data}.  The primary reasons were the lack of
suitable methods of analysis, on the one hand, and of suitable observational
data sets, on the other hand.  However, in the last decade a great deal of
progress has been made in nonlinear time-series analysis.

Seemingly erratic data can have a strong deterministic component in addition to
a stochastic or noisy background.  The goal of nonlinear signal processing is
to detect the presence of such a deterministic signal, and to infer information
about the dynamics that generates it, most often without {\sl any a priori}
knowledge thereof.  Such a feat is possible if the dynamics is sufficiently
low-dimensional which means that the signal is generated by a flow or a map
with only a few variables.  One can then hope to make a connection between
these variables and corresponding physical variables, and to obtain a physical
understanding of the pulsational behavior.  We refer the reader to some recent
reviews, \eg Weigend \& Gershenfeld (1994), hereafter WG94; Abarbanel \etal
(1993), hereafter ABST93; Casdagli \etal (1992).

In astronomical observations we are usually limited to measuring {\sl a single
variable}, namely the magnitude (and perhaps, the radial velocity).
Fortunately a very powerful tool has emerged from nonlinear science in the form
of embedding theorems.  Quite generally, in order to obtain information about
the dynamics it is not necessary to measure {\sl all} the phase-space
variables, but it is sufficient to know the temporal behavior of {\sl one} of
them.  In fact, it is enough to know the behavior of a single variable which is
a function of the phase-space variables, such as the luminosity or the radial
velocity.  To apply the theorems to an observational time-series it is
necessary however to sample the phenomenon at equally spaced time-intervals.
While this is easy to do under the controlled conditions of a laboratory
experiment, in long-term astronomical observations it is essentially
impossible, whether because of telescope scheduling constraints and politics,
sky obscuration or other causes.

\begfigwid0cm
\hskip-0.3truecm\psfig{figure=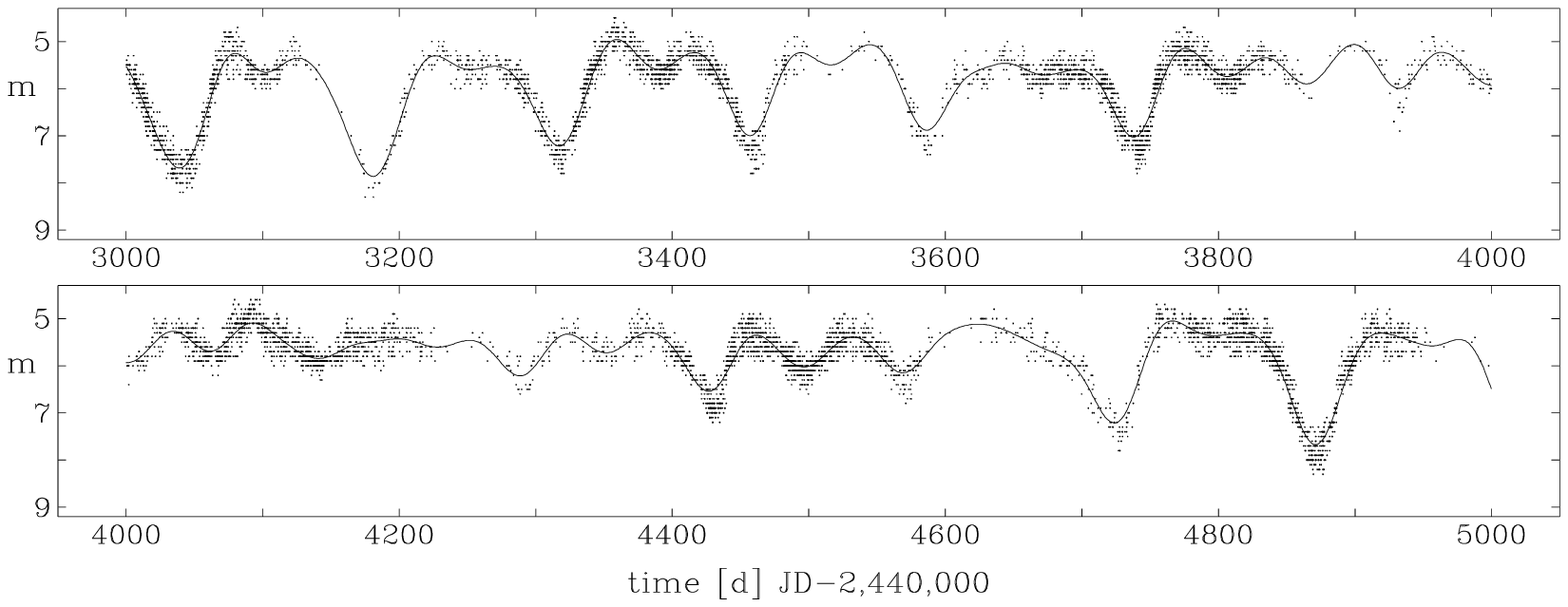,width=18.3truecm}
\figure{1} {Typical observed light curve segments for R~Scuti.
{\sl Dots:} the individual observations, {\sl line:}
the smoothed filtered  signal.}
\endfig

Nonlinear techniques are new in Astronomy, and, quite generally, most
astronomical observations have not been made with a nonlinear analysis in mind.
As a consequence the measured light-curve data are often not suitable for the
modern tools of analysis.  Rarely are enough pulsation cycles covered by the
observations to provide a 'typical' sample.  The sampling rate, \ie the number
of points per cycle, is often too small.  Large, and even regular gaps in the
data pose no problem (in contrast to the standard aliasing problems in linear
Fourier analyses), but data segments that are too short can make an analysis
impossible.  Finally, the signal to noise ratio is often too small.

Recently, the American Association of Variable Star Observers (AAVSO) has
completed a compilation of the observational data on some RV~Tau type stars.
Of these the R~Scuti light-curve is particularly long (last 31 years) and
relatively well sampled.  Furthermore this star has a large pulsational
amplitude, making the observational errors relatively small.  We have taken
advantage of the availability of the AAVSO data, and of the power of the global
flow reconstruction method (Serre, Koll\'ath \& Buchler 1995a, hereafter SKB),
to analyze its pulsations.  

In order to establish the reliability of the global flow reconstruction method
we have presented extensive tests in SKB.  These tests were performed on the
well studied known R\"ossler attractor which consists of 3 ODEs (\cf also Brown
1992).  Noise was also added to the R\"ossler data in order to simulate a more
realistic astronomical environment.  Our reason for choosing the R\"ossler
attractor for benchmark tests is that, superficially at least, it bears a
strong resemblance to the attractor that has been found in the pulsations of
W~Vir models (BK87; KB88).  In SKB it is shown that from the mere knowledge of
a relatively {\sl short} stretch of the temporal behavior of {\sl only one} of
the R\"ossler variables the global flow reconstruction technique correctly
recovers (a) the dimension of 3 for the R\"ossler system, and (b) the Lyapunov
exponents and fractal dimension of the attractor.  (We note that for such short
a data set a computation of the correlation dimension would not be able to come
up with a good estimation of the fractal dimension, and would not give any
information about number of ODEs that generate the R\"ossler band).

We have also applied the global flow reconstruction technique to the analysis
of the pulsations of one of these W~Vir models.  This numerical hydrodynamical
study which discretized the star into 60 mass shells thus corresponded to the
solution of 180 nonlinear coupled first order ODEs (180D space).  The flow
reconstruction shows that the W~Vir model pulsations are found to be generated
by a mere 3 dimensional flow.  The reader may appreciate that this dimensional
shrinkage from 180 to 3 is nontrivial, neither mathematically nor physically.
The physical implications of this result are discussed in Serre \etal (1995b).

In a Letter (Buchler, Serre, Koll\'ath, \& Mattei 1995, hereafter BSKM) we have
already described the highlights of our analysis of the lightcurve of R~Sct.
Here we provide not only a more thorough description of the tests that lead to
that conclusion, but we also present additional tests.  Furthermore, we discuss
the physical implications of this result.  In a companion paper (Koll\'ath,
Buchler, Serre \& Mattei 1995) we will communicate the analysis of AC~Herculis,
another star of the same type.

In \S2 we discuss the data preparation procedure.  In \S3 we present two
standard astronomical analyses, \viz a multi-periodic Fourier decomposition and
an autoregressive (AR) scheme, respectively.  We show that neither are
compatible with the observational data.  In \S4 we present the results of the
global flow reconstruction of the R~Sct data followed by a discussion in \S5.
A physical picture of the uncovered low dimensionality of the dynamics
is given in \S6.  In \S7 we conclude with an evaluation of the global
reconstruction method and its prospects for Astronomy.

\titlea{\ \ THE DATA PREPARATION}

An important problem arises in connection with the separation of the measured
data into 'signal' and 'noise'.  In order to extract information about the
dynamics which generates the light-curve, it is clear that one wants to
eliminate as much as possible observational noise, as well as noise that is
extraneous to the star.  Even then, however, there generally will still be a
remaining 'noise' that has its origin in convection and turbulence, and which,
strictly speaking, is part of the dynamics.  It is well known that turbulent
motions are usually very {\sl high}--dimensional behavior (they involve many
modes), and this therefore would seem to defeat \apriori\ the purpose and any
attempt to uncover any {\sl low}--dimensional behavior, unless the pulsation
can be separated into a large amplitude, low-dimensional 'true' pulsation and a
low amplitude, but high-dimensional jitter.  The success of our analysis seems
to indicate that in the case of the pulsating stars of interest to us here,
such a separation is indeed possible and that the high-dimensional jitter can
be eliminated together with the true noise, leaving the low-dimensional signal
relatively undisturbed.

\begfig0cm
\hskip-0.3truecm\psfig{figure=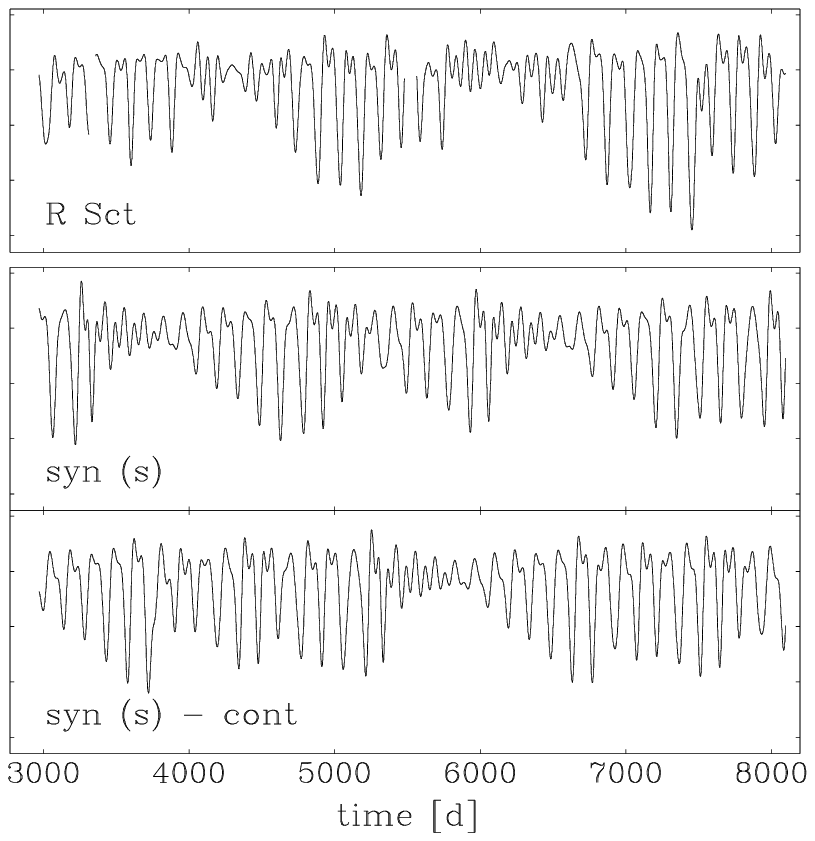,width=8.5truecm}
\figure{2} {{\sl Top:} The smoothed light curve of R~Scuti, Julian Day
2,438,000+; $s$ data set. {\sl Below:} two segments of the same synthetic
signal.
}
\endfig

\begfig0cm
\hskip-0.3cm\psfig{figure=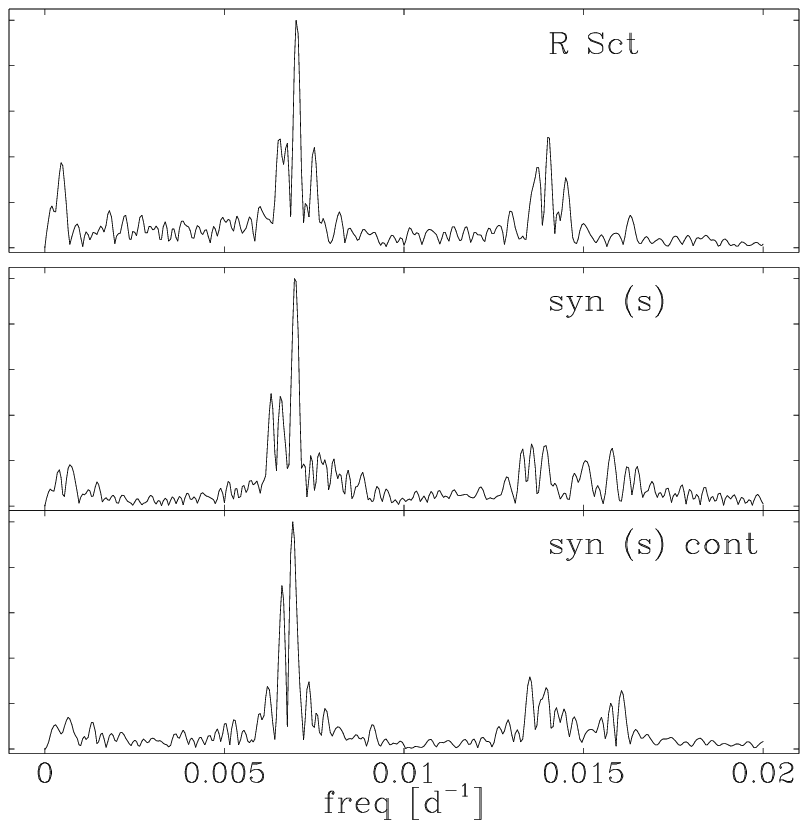,width=8.5truecm}
\figure{3} {Amplitude Fourier spectrum for the signals of Fig.~2.
}
\endfig

The smoothing and filtering of the data is therefore an important and delicate
part of any analysis of 'real' data.  Here, too little smoothing will not
suppress the high-dimensional component and preclude a successful search for
low-dimensional behavior.  On the other hand too much smoothing can severely
distort or even destroy the low-dimensional component.  Furthermore, certain
types of filtering (IFR) can increase the embedding dimension (but not reduce
it!, {\it q.v.\ }  ABST93).


The AAVSO observational data consist of visual estimates of the magnitude.  A
short typical sample of the raw observations (dots) is shown in Figure~1.  One
sees that the sampling can vary from excellent to poor.  A statistical analysis
of the data shows that the individual observations have a normal distribution
with $\sigma_s=0.2$ that is independent of magnitude, presumably because of
the logarithmic response of the human eye.  We have thus chosen to work with
the magnitude rather than with the luminosity.

\begfigwid0cm
\hskip-0.5truecm\psfig{figure=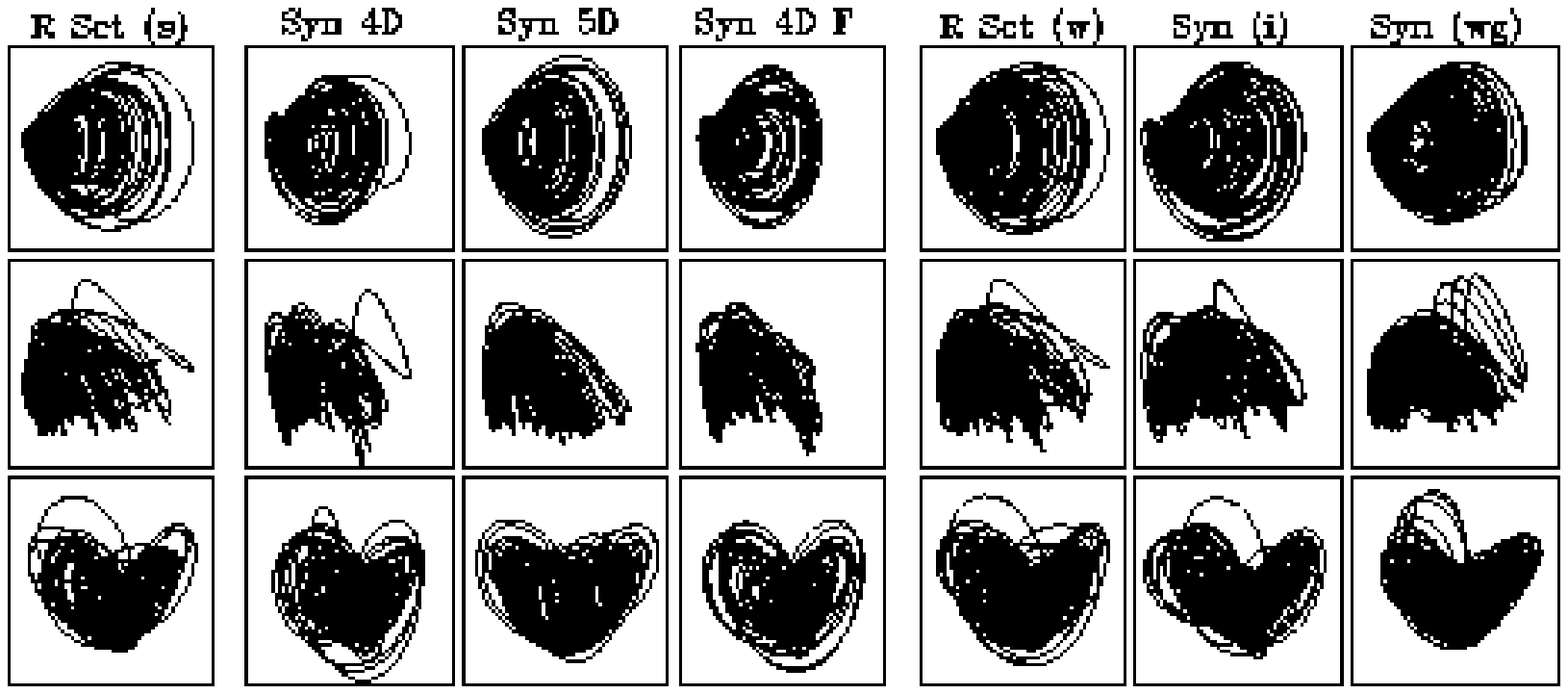,width=18.truecm} 
\figure{4} {Lowest 3 BK projections.
     {\sl Col.  1:} $s$ data set, {\sl 2--3:} synthetic signals from map
        (\cf text), 
     {\sl 4:} synthetic {\sl flow},
     {\sl 5:} $w$ data set, {\sl 6 \& 7:} synthetic signals from $i$ 
     and $w$ data sets. --
NOTE: the quality of this figure is crappy due to xxx.lanl.gov requirements}
\endfig

Our preprocessing of the raw observations proceeds as follows. First, we
average the data on 2.5 day intervals.  The precise value of the averaging
interval is not critical, as tests with 5 day averages have shown.  These
averaged data are then smoothed and interpolated with a cubic spline (Reinsch
1967) in which the only free parameter is the estimated noise level of the
input data ($\sigma_s$).  The final standard deviation of the magnitude
averages from the smoothed curve is thus determined by this quantity.  The last
step in the preprocessing of the data is a low-pass Fourier filtering with a
frequency cut-off of 0.02 c/d at 3dB.  We have also experimented with other
smoothing techniques, such as Savitzky-Golay filtering (\eg Press \etal 1992)
and Fourier down-sampling (Sauer in WG94), which however, if anything, have
been less satisfactory.  

The final result of the preparation process is a data set $\{g(t_n)\}$ that is
sampled at constant 1 day intervals.  In Fig.~1 we have shown the type of
smoothed curve that we fit to the raw observational data.  The fit misses
somewhat the deep light-curve minima because of the smoothing and filtering
that suppresses the high Fourier components.

In Figure~2, on top, we display the 5\th 000 day, middle subset of the smoothed
AAVSO data, which we label $s$.  In most of our analyses, \S4.1 - \S4.3, we
have used this $s$ set, first because for reasons of an economy of time, and
second because it appears more typical in the sense that it contains intervals
of both small and large amplitudes.  The whole ($w$) 30 year AAVSO data set, to
be analyzed in \S4.4 set is shown in Figure~8 in which for reference the
horizontal bar indicates the short ($s$) subset.  

The amplitude Fourier spectrum of the $s$ data set is exhibited on top of
Figure~3.  (We do not show the Fourier spectrum for the $w$ data set because it
is very similar, \cf also Koll\'ath 1990 who has Fourier analyzed all available
150 years observations).

In the nonlinear analysis of \S4 we shall find it convenient to use
Broomhead--King (1987) (BK) projections which project onto the eigenvectors of
the correlation matrix.  These projections are optimal because on the one hand
they are orthogonal and on the other hand they provide an optimal spread away
from the diagonal.  In the first column of Figure~4 we display plots of the
lowest three BK coordinates $\{ x_k\}$ for the $s$ data set.  The successive
rows show the ($x_2$ \vs $x_1$), ($x_3$ \vs $x_1$) and ($x_3$ \vs $x_2$)
projections.  In these and in the following BK plots the scaling of the axes is
$x_1$:$x_2$:$x_3$= 8:4:1.  One notes that the $w$ data set, displayed in column
5 looks very similar in BK projections.

Before we show our nonlinear analysis we will first apply two standard
astronomical techniques to the observational data.  Both are based on the
assumption that the signal is multi-periodic, and we show that neither of them
gives acceptable results.

\titlea{\ \ THE FAILURE OF STANDARD LINEAR ANALYSES}

\titleb{Fourier Decomposition}

The first question one may want to address is whether a multi-period Fourier
sum can give a good description of the data.  We have therefore constructed a
least-squares fit for a multi-periodic signal using, quite generously, the 35
most significant frequencies and their amplitudes and phases (of the $s$ set).
(We have chosen a 70 parameter fit for future comparison purposes,
because that will also be the number of parameters of the successful nonlinear
map to be described in \S4).  For consistency with the following nonlinear
analysis we have also used the smoothed signal for this analysis although the
smoothing has little effect on the Fourier fit and on the conclusions we reach
about its usefulness.

\begfigwid0cm
\hskip-0.5truecm\psfig{figure=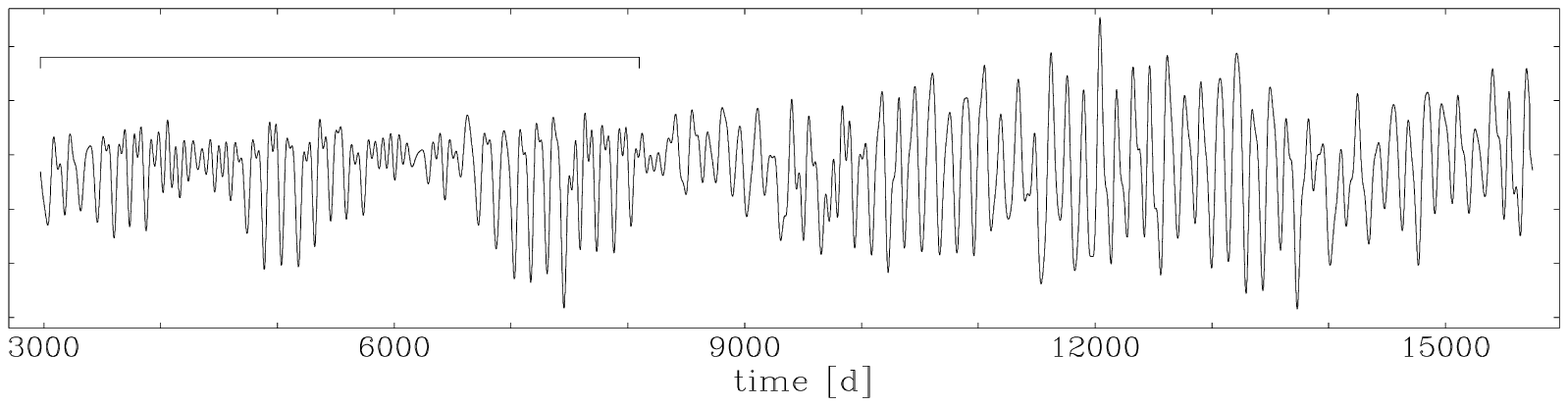,width=18.truecm}
\figure{5} {Multi-periodic synthetic signal fitted to the $s$ data set.}
\endfig

\begfigwid0cm \hskip-0.3truecm\psfig{figure=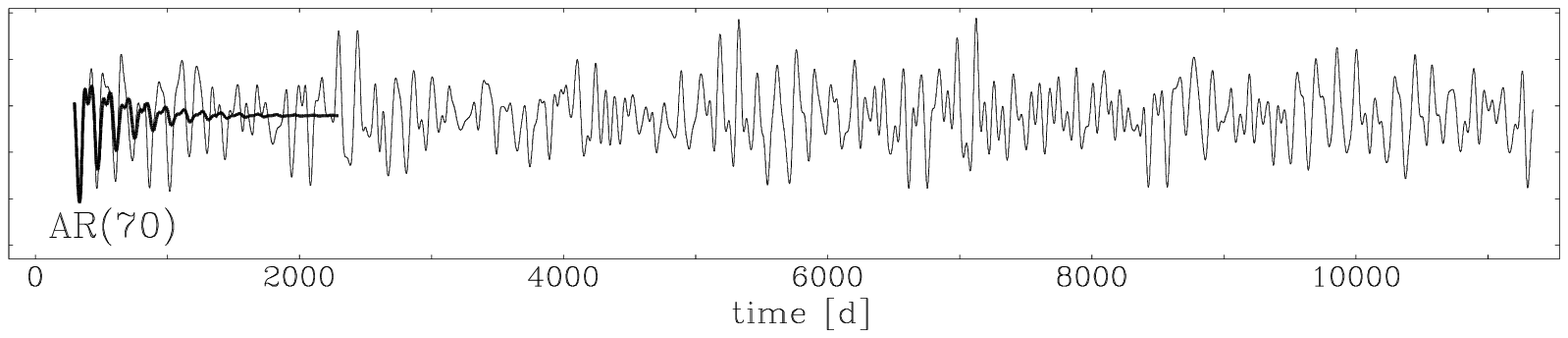,width=18.3truecm}
\figure{6} {A typical implementation of an autoregressive scheme.}
\endfig

The multi-periodic Fourier fit is shown in Figure~5.  A comparison with Fig.~2
shows that {\sl it is quite good over the data set which is delineated with an
overbar.}  However, beyond the range of the $s$ set it displays a behavior that
ostensibly is {\sl not compatible} with the observed R~Sct data.  This is even
more convincing when the light-curve data over the last 150 years (Koll\'ath
1990) are considered.  The Fourier expansion can thus serve as a good
interpolation, but it fails miserably as an extrapolation.  This should not
really come as a surprise because it takes many coefficients and very strict
phase-relationships to produce a signal with the kind of asymmetry that the
observed light-curve possesses.  In the extrapolated regime the phases act as
if they were random and thus produce the much more symmetric signal seen beyond
$t$=10\th 000 in Fig.~5 in which the characteristic RV~Tau type alternations of
deep and shallow minima disappear.  In contrast, as we shall see, the nonlinear
map not only fits the data set, but it also has predictive power in the sense
that the nature of the synthetic signals is very similar to the data set,
independently of the epoch.

\vskip 10pt

We also would like to point out that there is also a theoretical reason that
militates against a multi-periodic interpretation of the light curve.  The
Fourier spectrum shows a large number of frequency peaks above the noise level
that would have to be interpreted as modal frequencies or linear combinations
thereof.  There are however very few radial modes in the observed Fourier
spectral range.  Furthermore, the visual whole disk observations can detect
nonradial modes up to $\ell$=3 at best, and only the lowest few of these can be
expected to be vibrationally unstable.  There is thus a shortage of modes to
explain the spectrum as multi-periodic.

The question arises whether the Fourier spectrum changes over the 150 years
could be due to evolutionary changes in the sructure of the star.  The work of
Koll\'ath (1990) has shown that the envelopes of the Fourier spectra are
relatively steady for different sections of the available 150 years of data,
but the individual peak structure is not.  Even if, contrary to our previous
conclusion, one could interpret the peaks as due to a very large number of
excited modes, the observed frequency peaks in successive temporal segments do
not appear to move in a systematic fashion with time as one would expect from
an evolving star (which in addition would have to evolve much more rapidly than
evolution theory indicates).  A multi-periodic interpretation of the observed
Fourier spectrum is therefore essentially incompatible with pulsation and
evolution theories.

\vskip 10pt

\titleb{Autoregressive Analysis}

\vskip 10pt

We have also performed another common linear analysis, namely an AR fit which
assumes that the signal is produced by a linear auto-regressive process with
white noise (\eg Scargle 1981, WG94).  The physical picture behind this
approach is that of a set of linear modes which are stochastically driven.  

The thin line in Figure~6 represents an implementation of a 70 coefficient AR
process, with a noise intensity such as to reproduce the overall amplitude of
the observed fluctuations.  The signal of Fig.~6 is plotted on the same
vertical scale as the data set (Fig.~2).  {\sl The AR signal ostensibly again
bears little resemblance to the $s$ data set}.  In contrast to the Fourier fit
of the previous section, the AR extrapolation preserves some of the RV Tau-like
alternations of deep and shallow minima, but it lacks the asymmetry of the
observed light-curve.  Using as many as 500 coefficients does not help.  We
note in passing that using an ARMA instead of this AR does not improve the
signal either.  The thick curve in Figure~6 shows the extrapolation of the AR
process without noise, started on the data set.

We note that this AR analysis also dodges the question as to the mechanism
which could provide a stochastic driving of sufficient strength to produce the
observed irregularity.

The AR analysis again confirms that the dynamics that generates the signal is
strongly nonlinear.

\titlea{THE GLOBAL FLOW RECONSTRUCTION}

Our nonlinear analysis falls under the general category of {\sl global flow
reconstructions} (\eg ABST93) and the details of our approach are clearly
documented in SKB to which we need to refer the reader.  Space permits us to
present only the gist of it here.

We assume that the pulsations are generated by a flow of the form of $d{\bY}/dt
= {\bG}({\bY})$, where ${\bY}(t)$ traces out the trajectory of the system in
its $d$--dimensional {\sl effective phase-space}.  \Apriori we have no clue
about the function $G$, nor even its dimension $d$.  According to the embedding
theorem there then exists a variable ${\bX}$ in the $d_e$--dimensional {\sl
reconstruction-space} which satisfies a nonlinear equation (map) of the form

$$\quad\quad {\bX}^{n+1} = {\bF}[{\bX}^n],
\eqno(1)$$

\ni The set of $d_e$--dim vectors ${\bX}^n =\{g(t_n), g(t_n\ngth-\ngth\Delta),
g(t_n\ngth-\ngth2\Delta), \ldots , g(t_n\ngth-\ngth (d_e\ngth-\ngth 1)\Delta)
\}$ can be constructed {\sl from the observed scalar variable} $\{g(t_n)\}$, in
our case the star's magnitude, and the $\{t_n\}$ are equally spaced times.  The
time-delay $\Delta$ is an integer multiple of the spacing.  The $g$ variable is
assumed to be a smooth function of the effective phase-space variable $\bY$,
\viz $\{g(t_n)\}\equiv\{g({\bY}(t_n))\}$.  We will refer to $\bX^n$ as a {\sl
trajectory} because it maps out the temporal behavior of the system in the
$d_e$--dim Euclidian reconstruction-space.  (In \S4.3 we shall also reconstruct
a $d_e$ dimensional flow (eq.~2) instead of the map of eq.~1.).

The physical trajectory $\bY (t)$ is of course devoid of singularities and
intersections, but the transformation (diffeomorphism) from $\bY\in {\bf R^d}$
to $\bX\in {\bf R^{d_e}}$ can in general cause such ambiguities to appear in
the trajectory $\bX^n$.  However when the dimension $d_e$ is large enough all
cusps and intersections become resolved and the transformation is then called
an {\sl embedding}, and the reconstruction-space becomes an {\sl
embedding-space}.

The embedding theorem establishes a one-to-one correspondence between the
attractor points in the effective phase-space and the one reconstructed from
the embedding variable.  This is a very powerful result because some quantities
are preserved in the embedding.  It thus allows one to {\sl infer quantifiable
properties of an unknown underlying dynamics from the observations of a single
quantity}, namely the magnitude in our case.

In SKB we have shown how the reconstruction analysis yields both a lower and an
upper bound on the \apriori\ unknown dimension $d$ of the dynamics.  In the
case of R~Sct it will turn out that the lower and upper limits coincide which
will allow us to infer a definite dimension $d$ for the dynamics.

The best choice for the global map ${\bF}$ is of course unknown.  It is natural
to try a polynomial expansion.  Thus we assume ${\bF}({\bX}) = \sum_{k} {\bf
C}_k P_k({\bX})$, where the $P_k({\bX})$ represent polynomials of order up to
$p$ that are {\sl orthogonal on the data set}.  In SKB we discuss how the
polynomials are constructed and the coefficients are determined efficiently
with a singular value decomposition (SVD) method.

Once we have constructed a map $\bF$ we can generate {\sl synthetic}
trajectories or signals {\it ad libitum} by iterating the map (eq.~1) with some
seed value.  The nontrivial question then arises how to compare the chaotic
synthetic signals with each other and with the data set.  Standard statistical
comparison tests are not possible because of the small size of the data set.
We are therefore obliged to resort to more subjective tests.  The comparison of
the appearance of the synthetic signal to the original one is usually not
enough.  For example, as we have seen in \S3, multi-periodic signals can mimic
the observational light variations, but the difference is quite obvious from
the Fourier transform.  Also,  Fourier spectra have difficulty
distinguishing between signals with amplitudes modulations of different lengths
and depths.  Our experience is (SKB) that in addition to the signal itself and
its Fourier spectrum it is necessary also to compare projections of the
reconstructed phase-space trajectories.

In SKB it has already been discussed how and why small changes in the
smoothing, the polynomial order $p$, and the delay $\Delta$ can change the
synthetic trajectory between chaotic, periodic, or unstable.  In the case of
real data such as we consider here, there are additional parameters related to
the filtering, the treatment of gaps, and the interpolation.  {\it Prima
facie}, this sensitivity may appear very disturbing, but is to be expected.
Indeed, on recalls that even for the 1--D logistic map periodic cycles are
intimately close to chaotic solutions as the parameter of the map is varied
(O93).  It is therefore no surprise that one witnesses here a strong
sensitivity to the precise values of the coefficients of the polynomial map.

It is thus imperative to distinguish between {\sl the robustness of the
dynamics itself}, in particular the robustness of the structure and
dimensionality of system (1), and {\sl the extreme sensitivity of the integral
curves or trajectories of this dynamics to the parameters of the maps}.  The
important point is to ascertain that {\sl within a reasonably broad range of
the above mentioned parameters chaotic solutions exist} and that their
properties are robust.  By robust we mean that the synthetic signals satisfy
our three comparison criteria with the data set.  Furthermore, we require that
quantitative properties, such as the Lyapunov exponents and the fractal
dimensions (\eg O93) that we compute from the various synthetic data display a
certain invariance over the range of good maps (see also SKB).

It sometimes happens that a map is unstable in the sense that an iteration of
the map for the generation of a synthetic signal blows up.  This can occur when
the density of data points in regions of high divergence is insufficient to
provide a stable neighborhood, and small changes in the nonlinear coefficients
allow the orbit to be kicked away from the attractor.

Our methodology is to search for the best region of robust maps and synthetic
signals by exploring the parameter-space of the three main quantities: the
smoothing parameter, the time-delay and of course the embedding dimension.  The
order of polynomial expansion $p$ plays a lesser role as long as it is large
enough, but of course it is useful to find the smallest value.  The number of
SVD eigenvalues and their cut-off value (\qv\ SKB), as well as the locations
and widths of the gaps also come into play, but generally are less critical.  A
systematic search of the whole parameter-space is rather time-consuming so we
have used the strategy of searching with one parameter at a time.  After
several 'iteration steps' we can generally find the best regime in
parameter-space.  If this regime is broad enough we claim success.

\titleb{Reconstructions with the  $s$ data set}

The typical behavior of the error between the observed values and the values
predicted with the map as a function of the polynomial order $p$ has already
been shown in figure~1 of BSKM.  There is a definite convergence as $p$ is
increased.  More importantly the figure shows a clear levelling off beyond a
value of $d_e$=4.  While {\sl this suggests a minimum embedding dimension of
4}, by itself however it is not sufficient and we have to ascertain that the
4D synthetic signals have the same properties as the data set and that the maps
are robust.  

We note though immediately that, consistently with the behavior of the error
norm {\sl we have not been able to find a single good 3D map} with synthetic
trajectories that bear any resemblance to the R~Sct data set.  On the other
hand, our extensive tests with the R~Scuti show that the global flow
reconstruction provides excellent maps in 4D with the parameters $4\le \Delta
\le 7$, $p=4$, and $\sigma_s=0.1$.  It thus appears that the minimum embedding
dimension is indeed equal to 4 or greater.  This conclusion will be further
supported below by the fact that the fractal dimension of the attractor is also
greater than 3.

Two contiguous sections of one of our best representative 4D synthetic signals
($p=4$, $\Delta=6$, $\sigma_s=0.1$) are displayed in Figure~2 (rows 2--3).  The
amplitude Fourier spectra of these two segments are also depicted in Figure~3.

The lowest 3 BK projections of the observations and several different sets of
synthetic data are displayed in Figure~4.  Although the maps have been obtained
for various $\Delta$ and $d_e$, when we want to make comparisons we have to
construct the various BK plots with common parameters which throughout we have
chosen as $\Delta$=2 and $d_e$=8.  The 4D synthetic signal shown in col.~2 has
been constructed with $p$=4, $\Delta$=5, and the 5D signal in col.~3 with
$p$=3, $\Delta$=7.  For both the smoothing parameter is $\sigma_s$=0.10.

The synthetic signals themselves (Fig.~2), their Fourier spectrum
(Fig.~3) and their BK projections (Fig.~4) are very similar, and {\sl they are
independent of the embedding dimension}.  This strongly corroborates that {\sl
the minimum embedding dimension is indeed equal to 4}.

\vskip 10pt

Our analysis has a number of free parameters and we now address the
important question of the robustness of the results with respect to these
parameters.  It is the time-delay $\Delta$ and the smoothing parameter
$\sigma_s$ that play the most important role in the analysis (SKB).  A brief
comparison of the effects of these parameters is therefore presented in
Figure~7 for the lowest BK coordinate projection.  The middle row corresponds
to a value of the smoothing parameter $\sigma_s$=0.10 for which we obtain the
best maps.  In SKB we have found that a value of $\sigma_s$ is equal to the
actual noise level gives the best maps.  Here we do not know the exact value of
the average observational error, but the error for the averaged magnitudes is
in the range $\sigma_s$=0.04--0.20 depending on the number of observations in
the 2.5 day bins. The optimal smoothing level is thus in agreement with the
level of the observational noise.  We note that good maps are obtained for a
range of $\Delta$ values from 4 to 10 (except for a limit cycle at $\Delta$=9).

{\sl A priori} we had not idea what values to choose for $\Delta$, or even
whether any values could be found for which one obtains a good map that
captures the dynamics.  All we could expect is that for small values the map
would be close to linear, but since the trajectory is then collapsed to the
diagonal noise would overwhelm the reconstruction.  On the other hand for large
$\Delta$ the map could become so nonlinear that perhaps a polynomial
representation would be inadequate.  Our results show that indeed there exists
a region of $\Delta$ in which good maps can be obtained.

Fig.~4 also demonstrates that with over-smoothed data ($\sigma_s$=0.12) the
maps do not give complex enough synthetic curves.  The plots clearly indicate
some multi-periodic (n--torus) behavior with distorted Lissajous-like curves
for lower time-delays.  In contrast, the slightly under-smoothed light-curve
($\sigma_s$=0.09) shows that chaotic signals are possible, but that the maps
are more unstable (we cannot iterate for more than some 20--30 thousands of
iterations, at most).

The low pass filtering following the spline smoothing helps in the construction
of good robust maps, presumably because a reduction of the high frequency power
decreases the nonlinearity of the signal.  We have to note that with larger
$\sigma_s$ in the spline smoothing we can obtain relatively good results
without the Fourier filtering.  However, we obtain the best results with
combined spline smoothing and Fourier filtering.  While a purist might object
to this distortion of the signal and hence of the attractor, we think that we
merely remove sharp features which do {\sl not} affect the essence of the
attractor nor the physical interpretation that we will give of the dynamics.

We have also tested the sensitivity to sampling rate.  The results remain
relatively good with a time-step of 2 days, but an even lower sampling rate
destroys the similarity between the light-curve and the synthetic signals.  We
attribute this breakdown to the polynomial map which may not be able to handle
the increased nonlinearity.

There is one more parameter controlling the reconstruction of the map with the
SVD procedure (Press \etal 1992), namely the eigenvalue cut-off $\omega_c$.
The number of combinations of monomials goes up very quickly with $p$ and $d_e$
as $C^p_{d_e+p}$ (\eg Casdagli 1992) and their independence over the data set
decreases rapidly, as indicated by small SVD eigenvalues.  The purpose and
power of the SVD method are that by using such a cut-off, typically at machine
precision (10$^{-14}$), one automatically selects the most important
combinations of monomials, and one rejects the others which give numerical
problems in Gram-Schmidt polynomial constructions or in QR decompositions.  The
cut-off $\omega_c$ is here defined as the ratio of smallest to largest SVD
eigenvalue.  For most calculations, as long as we keep $p$ small, we can use
all the meaningful eigenvalues for the reconstruction (For example in 4D with
$p$ = 4 there are 70 eigenvalues, all of which are above the cut-off).
Actually, the synthetic signals generally are insensitive to $\omega_c$ as long
as $\omega_c<10^{-10}$, and as long as $\omega_c$ does not exceed ten times the
ratio of the smallest to the largest eigenvalue.  For larger values of
$\omega_c$ the map loses flexibility ('under-fits') and we usually obtain limit
cycles and fixed points.

\begfigwid0cm
\hskip-0.5truecm\psfig{figure=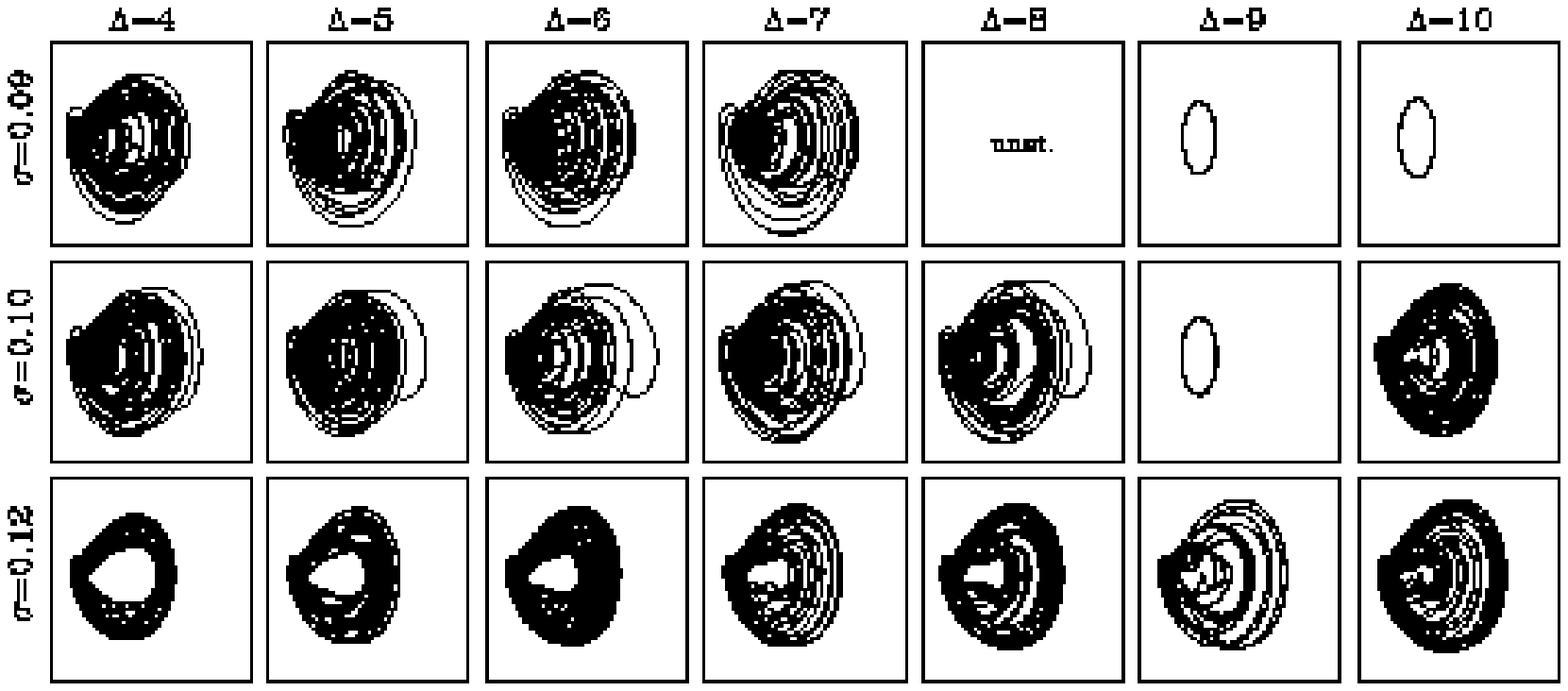,width=18.truecm}
\figure{7} {Lowest BK projections (4D maps) for different values of
$\Delta$ and $\sigma_s$.
 --
NOTE: the quality of this figure is crappy due to xxx.lanl.gov requirements}
\endfig

In BSKM we had inserted a small gap where we had estimated that the data were
not good enough to give a reliable interpolation.  Further tests have shown
that the reconstruction is not very much affected by the presence of absence of
these gaps.

\begtabfull
\tabcap{1} {Reconstructed Lyapunov exponents [d$^{-1}$] and Lyapunov dimension}
\halign{#\hfil&&\quad#\hfil\cr
\noalign{\hrule\medskip}
$d_e$ \hf &
$\Delta$ \hf &
$p$\hf &
\hf $\lambda_1$ \hf &
\hf $\lambda_3$ \hf &
\hf $\lambda_4$ \hf &
${\rm d}_{L}$ &
note\cr
\noalign{\smallskip\hrule\smallskip}
4&  4 & 4 & 0.0019 & --0.0016 & --0.0061 & 3.05 &\ $^{[0]}$ \cr
4&  5 & 4 & 0.0017 & --0.0014 & --0.0054 & 3.06 &\ $^{[0]}$\cr
4&  6 & 4 & 0.0019 & --0.0009 & --0.0051 & 3.19 &\ $^{[0]}$\cr
4&  7 & 4 & 0.0020 & --0.0011 & --0.0052 & 3.18 &\ $^{[0]}$\cr
4&  8 & 4 & 0.0014 & --0.0010 & --0.0049 & 3.07 &\ $^{[0]}$\cr
5&  7 & 3 & 0.0016 & --0.0005 & --0.0041 & 3.27 &\ $^{[0]}$\cr
6&  8 & 3 & 0.0022 & --0.0003 & --0.0018 & 3.52 &\ $^{[0]}$\cr
\noalign{\smallskip\hrule\smallskip}
4&  6 & 5 &0.0013 & --0.0004 & --0.0036 & 3.26 &\ $^{[1]}$ \cr
4&  6 & 5 &0.0015 & --0.0011 & --0.0030 & 3.15 &\ $^{[2]}$ \cr
\noalign{\smallskip\hrule\smallskip}
4&  6 & 5 &0.0015 & --0.0013 & --0.0039 & 3.04 &\ $^{[3]}$ \cr
\noalign{\smallskip\hrule\smallskip}
\noalign{\ni{ 
[0] $s$ set, 
[1] $wg$ set,
[2] $w$ set,
[3] $i$ set}}
\noalign{\smallskip\hrule}
}
\endtab

The Gram-Schmidt construction that has been used in BSKM to generate the
polynomials and the map is much less robust and is more prone to round-off
errors than the SVD approach we have used subsequently in SKB and here.
Furthermore, our search for regions of good maps was somewhat less thorough in
BSKM.  The synthetic signals here are therefore slightly different from BSKM
although our conclusions are immune to this kind of detail.

\titleb{Lyapunov exponents and fractal dimension}

The important quantitative characteristics of chaotic systems (such as Lyapunov
exponents and fractal dimension, \eg O93) cannot be obtained directly from the
short observational data, but such calculations are possible for synthetic
signals because these can be computed for several hundred thousand points.  In
order to give a indication of the robustness of the Lyapunov exponents and the
Lyapunov dimension we show in Table~1 their variation with embedding delay.
The errors are about 10$^{-4}$ on the exponents, and about 0.02 on $d_{_{L}}$.
The positive Lyapunov exponent is seen to be quite robust.  The negative ones
show more scatter, but that is to be expected (ABST93).

For all the chaotic synthetic trajectories the second Lyapunov exponent is {\sl
always} smaller than 10$^{-4}$ \th ($\lambda_2$ has therefore been omitted from
Table~1.  This fact turns out to be very important.  Indeed, had we
constructed a flow rather than a map one of the exponents would have been
exactly zero.  Such a flow, on the other hand, when sampled at discrete
time-intervals becomes equivalent to a map.  When the sampling time is
sufficiently short, as it is for R~Sct, the Lyapunov exponents of the map and
the flow should be very close to each other.  {\sl The smallness of $\lambda_2$
thus provides a very strong independent check on our analysis and on the basic
assumption that there is a low-dimensional flow that determines the behavior of
the star}.

We find that there is perhaps a tendency for the Lyapunov dimension to increase
(and to deteriorate, if we make a parallel with SKB) as we increase $\Delta$,
presumably caused by the need for an increase in the map's nonlinearity.
There is a similar behavior with embedding dimension.  This deterioration is
almost certainly related to the decrease in the density of points in higher
dimensional embedding-spaces.

The positive sign of at least one Lyapunov-exponent (\ie $\lambda_1>0$)
guarantees that the attractor is chaotic.  The inequality,
$\lambda_1+\lambda_2+\lambda_3>0$, which is very solidly satisfied, implies
that the {\sl fractal Lyapunov dimension $d_{_{L}}$ is greater than} 3 (O93).
In no case have we ever found 2 positive Lyapunov exponents (hyper-chaos).

Table~1 also reveals the important fact that {\sl the Lyapunov exponents and
dimension are essentially independent of the embedding dimension $d_e$}, as
long as $d_e\geq 4$.  (The slight increase of $d_L$ with $d_e$ is most likely
due to the deterioration of the method because of a decrease in point density
with embedding dimension).  If a high-dimensional dynamics were at the origin
of the R~Sct pulsations $d_L$ would have increased essentially as fast as
$d_e$.

We have further calculated the correlation dimension $d_2$ (O93).  Because an
accurate computation of $d_2$ necessitates extremely large data sets and
because the correlation dimension can be contaminated by systematic deviations,
we estimate our error on $d_2$ to be about an order of magnitude greater than
for $d_{_{L}}$.  The correlation dimension is also known to underestimate
sometimes the fractal dimension (see \eg Gouesbet 1991).  This could be the
reason that our estimates give $d_2$ = 2.8--2.9, \ie values smaller than 3.
These values are however compatible with the theoretical ordering $d_2 \leq
d_1\leq d_0$ \th\th (O93), assuming the Kaplan-Yorke conjecture, \ie $d_{_{L}}
= d_1$ to hold.

The embedding theorem involves the box-counting dimension $d_0$.  We have not
computed $d_0$, but from the difference between $d_{_{L}}$ and $d_{_{2}}$ we
have reasons to believe that all three dimensions are close to each other.  In
fact, according to the embedding theorem, to make a difference, $d_0$ would
have to be greater than 3.5.  From the fractal dimensions one therefore infers
an upper limit $d_e\leq 7$.  This seems to be a very generous upper limit,
because our analysis has shown that we can get excellent embeddings already in
$d_e$=4.  We caution though that we cannot be entirely sure that we have
avoided all trajectory singularities in 4D.

\titleb{Flows}

Up to this point we have searched for embeddings with {\sl maps}.  It may
therefore be objected that while we have been able to construct nonlinear maps
from the data set, this does not guarantee that there exists a corresponding
{\sl flow} in the effective phase-space (although the systematic smallness of
the second Lyapunov exponent has already given a strong hint that a flow
exists).

It is therefore important to test whether an embedding flow can be constructed
directly from the data (SKB), \ie whether we can find a nonlinear function
$\bH$ such that 

$$d\bZ/dt=\bH(\bZ),
\eqno(2)$$

\ni where $\bZ(t)$ is such that $\bZ(t_n)=\bX^n$.  Since we only know the
function $\bZ$ on a discrete set of points we approximate the derivative by a
discrete Adams--Moulton scheme (as suggested by Brown, Rulkov and Tracy 1994).

We find that embedding flows exist for almost all the parameters for which maps
exist, and that their synthetic trajectories have a very similar behavior.  (We
generate the synthetic signals by integrating the system by a fourth order
Runge-Kutta scheme with a fine time-step $\delta t=0.001 $).  However, the
flows tend to be somewhat less stable than the maps, \ie the integration
typically blows up after some ten thousands days.  Column~4 of Fig.~4 shows the
BK projections for the best flow.

\begfigwid0cm 
\hskip-0.3truecm\psfig{figure=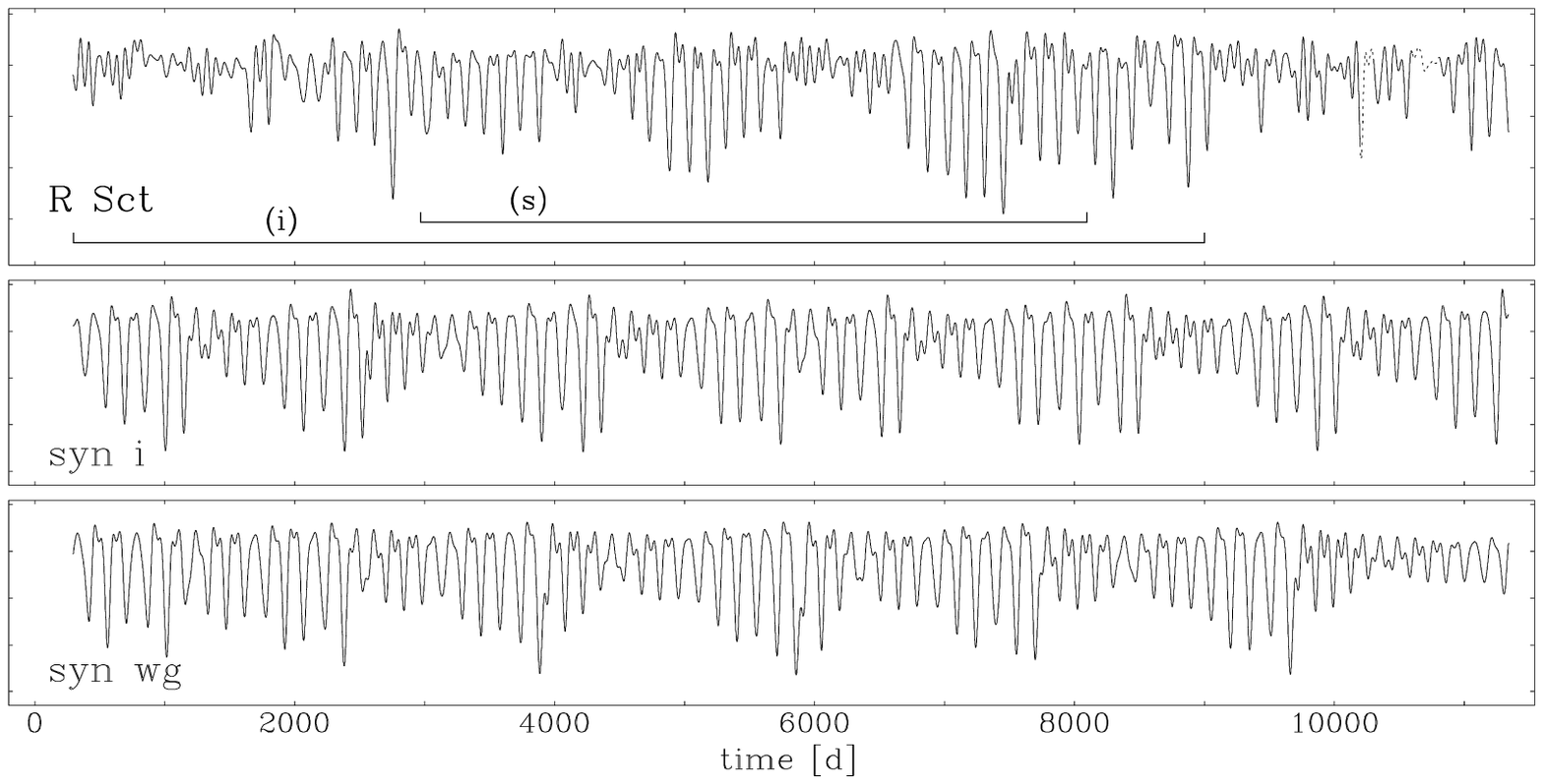,width=18.3truecm}
\figure{8} {{\sl Top:} The whole ($w$) smoothed AAVSO light curve of R~Scuti, 
Julian Day 2,438,000+. {\sl Below:} synthetic
signals from $i$ and $wg$ data sets, respectively.
}
\endfig

\titleb{Other tests}

Suppose we create a noisy, multi-periodic signal with a Fourier power-spectrum
that is very similar to the observational signal, but with different
phase-relationships.  Can we fool the global reconstruction. method into
believing that we have presented it with a low dimensional nonlinear flow?  We
have thus produced an artificial signal with the most important 35 frequencies
and amplitudes of the observational data, but have then randomized the phases.
Low dimensional polynomial maps constructed from this artificial data set are
incapable of producing synthetic signals with similar appearance!  From a
nonlinear dynamics point of view of course this is not surprising because the
artificial signal is multi-periodic and thus needs an embedding-space of a
dimension at least equal to 2$\times$ 35.

\vskip 10pt

The question also arises whether one might, erroneously, extract a
low-dimensional chaotic dynamics from a fully stochastic data set.  In theory
this is not possible, but it cannot be ruled out that for a real, noisy data
set one might find parameter combinations for which the maps give
low-dimensional chaotic synthetic signals.  However we deem it somewhat
unlikely that there would exist a {\sl range} of parameters in which a robust
map could be found.  As a test we have constructed a further artificial
stochastic signal by filtering Gaussian white noise with the envelope of the
Fourier transform of the R~Scuti data.  Again we could not make a good
reconstruction of these data.  The synthetic signals were always limit cycles
which represented some average of the trajectory (\cf also SKB).

\vskip 10pt

\titleb{Reconstruction with longer data sets}

In the preceding analyses we have only used a subsection ($s$ data set)
comprising the middle half of the AAVSO observations (\cf Fig.~2).  This
restriction to a subset of the whole data has been convenient because of
running-time economy for the large number of tests that we have made.  The
choice of this particular stretch of data has been dictated by its typical
nature with both large and small amplitudes.

We now turn to an analysis of the whole AAVSO data set which is shown on top of
Fig.~8.  We consider two slightly different versions of smoothed data sets,
one without gaps ($w$ set) and one ($wg$ set) with two gaps located at (10\th
200 $< t <$ 10\th 300) and (10\th 600 $< t <$ 10\th 800) where the
observational data are very sparse and the light-curve is poorly defined (the
dotted part of the curve represents the interpolation through the gaps).  For
reference, the horizontal bars in Fig.~8 denote the lengths of the $s$ and $i$
data sets.

The BK projections of the $w$ set are displayed in col.~5 of Fig.~4.  They are
seen to be hardly distinguishable from those of the $s$ data set, even though
the light-curve itself appears somewhat different qualitatively.  The envelopes
of the Fourier spectra of the $s$ and $w$ data sets are essentially the same as
well.

In Fig.~8 we exhibit the best 4D synthetic signal 'syn $wg$', obtained with $p$
= 4, $\Delta$ = 6 and $\sigma_s$ = 0.13.  In column~7 of Fig.~4 we display its
BK projections.

Even though these maps have been constructed with the $wg$ data set the
synthetic signals bear a stronger resemblance to the $s$ data set, as Fig.~8
shows.  In fact, the agreement with the $wg$ data set itself is perhaps not as
satisfactory as one would like.  The low amplitude, relatively more erratic
behavior in the last 1\th 500 days is not well mimicked by the synthetic
signals, although their Fourier spectra and the BK projections are quite
similar to those of the $wg$ set.

\begfigwid0cm
\hskip-0.65truecm\psfig{figure=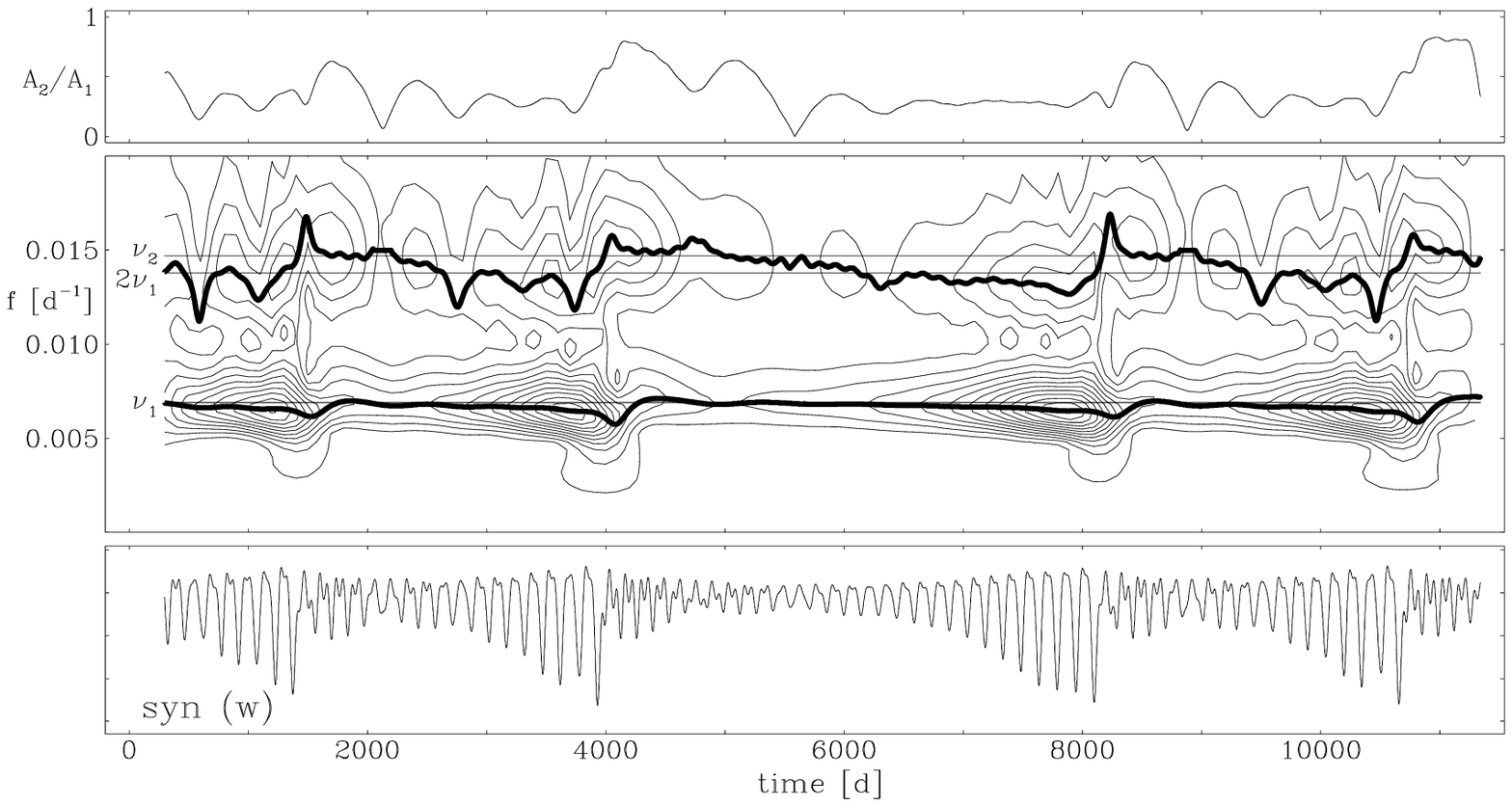,width=18.3truecm}
\figure{9} {Wavelet analysis. {\it Top:} Amplitude ratio, {\it middle:}
    Amplitude contours, {\it bottom:} $w$ synthetic signal from $w$ set.} 
\endfig

In view of the slight, but definite deterioration of the reconstructions with
the longer data set we have therefore also considered a number of sets of
intermediate lengths.  In Fig.~8 we have delineated one such set ($i$), which
comprises all the data with $t \le 9\th 000$.  The best 4D synthetic signal
that we obtain from this $i$ set is displayed in row~2 of Fig.~8 ('syn $i$')
and its BK projections are shown in column~6 of Fig.~4.  The best parameters
are the same as for the $wg$ set, \viz $p$ = 4, $\Delta$ = 6 and $\sigma_s$ =
0.13.  The synthetic signal bears a good resemblance to the $i$ set, but the
maps again seem somewhat reluctant to give long stretches of low amplitude
behavior such as occur for $t<1\th 500$ in the $i$ data set.  From Table~1 we
see however that the Lyapunov exponents and Lyapunov dimensions are the same as
for the synthetic signals generated from the $s$ set.

Generally we find that extending the $s$ data set to the left (earlier data)
has little effect on the maps and on the shape of their synthetic signals.  On
the other hand, when we add later data (on the right), the maps become more
sensitive to the parameters and to the location of the gaps.  Furthermore, even
the best maps do not reproduce quite as well as one would like the low
amplitude behavior seen in the AAVSO data for $t>9\th 000$ in row~3 of Fig.~8.
Below we shall return to a discussion of the possible reasons for these
discrepancies.

Another feature of the reconstructions from the longer data sets is the
enhanced sensitivity of the synthetic signals to small changes in the
parameters, and thus in the maps.  For example, the synthetic signal ('$wg$')
shown in Fig.~8 has been generated by a map that uses only 58 of the total of
70 SVD eigenvalues that occur with $d_e$ = 4 and $p$ = 4.  The use of all 70
eigenvalues causes a change of the order of 0.001\% in the map over the
observed data.  (We compare here the average values of the daily predictions of
the two maps calculated on the data set).  On the other hand, if we compute
similarly the daily predictions on a synthetic data set, the differences in
prediction between the two maps jump to 10--20\%.  This shows that the two maps
are very similar on the attractor, but that they become quite different in its
vicinity.  The use of more than 58 eigenvalues results in more regular
synthetic signals, some of which can even be limit cycles.

A similar sensitivity is observed with respect to the locations and lengths of
the gaps that we introduce in the data sets.  For the ungapped $w$ set, for
example, we obtain a chaotic signal with an amplitude modulation time-scale
that, compared to 'syn $wg$', is almost a factor of 2 longer (bottom,
Figure~9).  The pulsation amplitude grows much more slowly from small to large
with a 'more regular' appearance; then, in a briefer irregular phase the signal
collapses to small amplitude.

This behavior gives us a clue as to the origin of the sensitivity.  Indeed it
is very reminiscent of what can occur in the proximity of a Shilnikov-type {\sl
homoclinic connection} (\eg Glendenning \& Tresser 1985).  It then suggests
that we look for the fixed points of the maps and their linear stability roots.
From the polynomial expressions of the maps they are easy to locate.  In all
the maps we find a fixed point located at the center of the attractor, and its
two pairs of linear stability roots are always spiral.  In Table~2 we display
these two roots, $\pm 2i\pi\nu + \rho$, for the best synthetic signals for the
$s$, $i$, $wg$ and $w$ data sets.  First one notes that the two frequencies are
almost invariant, and that they correspond closely to the dominant frequency
peaks of the Fourier spectrum (Fig.~3).

\begtabfull
\tabcap{2} {Linear stability roots of fixed point of the maps}
\halign{#\hfil&&\quad#\hfil\cr
\noalign{\hrule\medskip}
\hf \quad data set \hf & $\hf \nu_1$ \hf & $\hf \rho_1$ \hf & \quad & 
$\hf \nu_2$ \hf & $\hf \rho_2$ \hf & \th
\cr
\noalign{\smallskip\hrule\smallskip}
\hf $s$ \hf  & 0.0068 &  0.0044 & & 0.0145 & --0.0062 \cr
\hf $i$ \hf  & 0.0069 &  0.0009 & & 0.0147 & --0.0019 \cr
\hf $wg$\hf  & 0.0069 &  0.0004 & & 0.0147 & --0.0013 \cr
\hf $w$ \hf  & 0.0069 &  0.0001 & & 0.0147 & --0.0012 \cr
\noalign{\smallskip\hrule}
}
\endtab

There is a strong correlation between these roots and the behavior of the
corresponding synthetic signals.  Consider first the $w$ signal (shown on the
bottom of Fig.~9) and its roots.  We interpret the signal as that of a
trajectory that expands slowly ($\nu_1/\rho_1 = 70$) along the mildly unstable
spiral manifold with the corresponding {\sl lower frequency} $\nu_1$ of 0.0069
c/d dominating.  Then this trajectory transits to the strongly attracting
($\nu_1/|\rho_2| = 6$) spiral manifold.  Here the higher frequency dominates
the pulsation.  This second phase is shorter consistently with the larger ratio
$\rho_2/\rho_1 = 12$.  

This interpretation is corroborated by a wavelet analysis of the $w$ synthetic
signal.  For details of this analysis we refer to Koll\'ath \& Szeidl (1993).
In Fig.~9 (middle) depicts the amplitude contours in a frequency -- time space.
The thick line represents the instantaneous frequencies $f_{inst, k}(b\ngth
=\ngth t) = \partial\phi(f_k,b)/\partial b$, where we have chosen $f_1=0.007$
and $f_2=0.015$, respectively, i.e. the dominant frequency peaks in the Fourier
spectrum (the precise values of $f_k$ are not critical because of the flat
behavior of the phases in the vicinity of these frequencies).  On top of Fig.~9
we record the ratio of the two wavelet amplitudes at $f_{inst, k=1,2}$.  The
shift back and forth between the harmonic $2 \nu_1$ and $\nu_2$ is nicely
brought to light by the wavelet analysis.  For example, in the expanding
stretch from $t\approx$ 2\th 500 to 4\th 000 the high frequency component of
the wavelet is dominated by the harmonic $2 f_1 \approx 2\nu_1 = 0.138$, but
during the subsequent attracting phase, from $t\approx$ 4\th 000 to 5\th 700
the high frequency component shifts to the second frequency $f_2\approx\nu_2=
0.0147$.  (The thin horizontal lines in Fig.~9 are drawn at these two
frequencies.)

The behavior of the other synthetic signals and the values of their stability
roots further corroborate this scenario.  Thus, for example, for the $s$
synthetic signal, the roots have a ratio $\rho_2/\rho_1 \approx 1.4$, and, in
addition, compared to the frequencies, the growth rates are now by a factor of
5 larger.  This is the reason for the more rapid amplitude variations of the
$s$ synthetic signal (Fig.~2) in which the expanding and contracting phases are
therefore also closer in length.  Furthermore, Table~2 shows that the real part
of the first root ($\rho_1$) is very sensitive.  It is thus not astonishing
that for some parameter values $\rho_1$ becomes negative and the synthetic
signal becomes a fixed point.  For positive $\rho_1$ the flow can also produce
limit cycles instead of chaotic trajectories (\eg Glendenning \& Tresser 1985)
as Fig.~7 indicates.

The evidence is seen to be strong that there exists a nearby homoclinic
connection in the maps (and, by inference, in R Sct itself) which can explain
the sensitivity of the long-term amplitude modulation to the parameters.  In
further support of this scenario we recall that the Lyapunov exponents and
Lyapunov dimension of the synthetic trajectories are essentially invariant,
suggesting that {\sl the dynamics itself is quite robust despite the extreme
sensitivity of the synthetic signals.}

\vskip 10pt

We now return to possible reasons for the deterioration of the synthetic
signals when, instead of the $s$ subset, the whole AAVSO data set is used.

As already mentioned Fig.~8 indicates that compared to the $s$ set there seems
to be a qualitatively somewhat different behavior for $t>9\th 000$, with
smaller amplitudes and a little more irregularity, and similarly perhaps for
$t<2\th 000$ as well.  In particular, one notes that outside the $s$ set there
are very few deep minima.  Extending the data set beyond the $s$ set therefore
provides no additional information about that part of phase-space.

We believe that it is the large relative noise level in the low amplitude
behavior (the observational error is about constant $\approx$ 0.1--0.2 mag
throughout the data set) that affects our ability to construct maps that
reproduce the long low amplitude modulations.  Evidence for the nefarious
effects of large noise intensity has already been presented in SKB: (a) when
large levels of noise were added to a chaotic signal the reconstruction often
led to limit cycles which represented merely an average behavior of the
pulsations, and (b) in an artificial signal that was composed of a limit cycle
plus colored noise the reconstruction ignored again the stochastic component
and recovered the average cycle.  In our situation here, instead of a limit
cycle, the 'average' behavior is chaotic.  In addition this average chaotic
behavior is dominated by the large amplitude pulsations that occur in range of
the $s$ set only.  This can be the reason why the best synthetic signals
systematically tend to have an appearance so similar to the $s$ data set.

We need to mention an alternative, less likely, but also less comfortable
explanation which might be found in the tacitly assumed constancy of the
stellar parameters and of the map.  Evolution time-scales for RV~Tau stars have
been estimated to be of the order of hundreds or thousands of years (Gingold
1974).  Could it be that the tiny structural changes in the evolving star cause
large changes in the nature of the trajectory in view of the strong sensitivity
of the signal to small changes in the map parameters?  We therefore caution
that this constancy is an assumption.  However it is a necessary working
hypothesis as it would obviously be impossible to gain much useful information
from such stars if the parameters were changing over a time-interval of the
length necessary to sample the attractor.

Notwithstanding the difficulty of the reconstruction to reproduce synthetic
signals with sufficiently long pulsational lulls it needs to be stressed again
though that the Lyapunov exponents and Lyapunov dimensions are robust,
suggesting that the reconstructed dynamics itself is robust, and close to the
actual dynamics of R~Sct.

\titleb{Spectral Analysis}

In addition to the Fourier spectra that we have already presented, we have also
computed a 70 parameter MEM spectrum.  The MEM spectrum of the data has just
two peaks and represents essentially an envelope of the Fourier spectrum
although the peaks are somewhat broader.

We now address briefly the connection between the Fourier spectrum and the map.
From our best 4D map we can construct a pseudo-MEM (amplitude) spectrum,

$$A(f )=\Biggl\vert {1\over 1-\sum_{k=1,4} a_k e^{-2i\pi\th
((k-1)\Delta + 1)\th f}}\Biggr\vert 
\eqno(3)$$

\ni (eq.~7, WG94), where the $a_k$ represent the {\sl linear} coefficients of
the 4D map around its fixed point.  This 4 parameter pseudo-MEM spectrum again
has two peaks with about the same widths as the envelope of the Fourier
spectrum, and is actually better than the 70 parameter MEM spectrum itself.
The locations of the two peaks of the pseudo-MEM spectrum coincide almost
exactly with the two frequencies of the linear stability analysis, $\nu_1$ and
$\nu_2$ of Table~2.  (It is interesting to point out that a comparable 4
parameter MEM spectrum misses considerably {\sl both} peaks).  The pseudo-MEM
low frequency peak at 0.0069 c/d is also very close to its counterpart in the
Fourier spectrum of R~Sct, \ie Fig.~3.  The high frequency peak, on the other
hand, occurs at $\approx$ 0.0147 c/d, which is somewhat higher than it is in
the Fourier spectrum.  This may not be a real discrepancy, and is likely to be
related to the unsteadiness of the signal.  Indeed, a comparison of the Fourier
spectra of two sections of the same synthetic signal in Fig.~3, middle and
bottom, shows a broad, unsteady structure at and beyond the first harmonic.
This is caused by the continual switching back and forth between the harmonic
of the first frequency ($2 \times$ 0.0069) and the second frequency $0.00147$
c/d, as uncovered by the wavelet analysis.

The pseudo-MEM spectrum is well defined only when there exists a corresponding
stationary AR process, which is not the case here.  When we iterate the linear
part of the map the amplitude of the lower frequency oscillation increases
without limit, and at the same time the higher frequency oscillation evanesces
(the rates of growing and decaying is given by the linear stability roots in
Table~2).  The nonlinear part of the map is needed to limit the amplitude and
to transfer power to the higher frequency.  This implies that the amplitude
ratio of the peaks in the pseudo-MEM spectrum has no physical meaning.

\titlea{DISCUSSION}

We have presented evidence that the pulsations of R~Scuti are {\sl not}
multi-periodic and that they are {\sl not} generated by a linear stochastic
process either, a process which, in any case would be hard to justify on
physical grounds.  Rather our global reconstruction has shown that the dynamics
of R~Sct, while noisy, is nonlinear and low-dimensional instead.  This result
is relatively robust with respect to the most important free parameters of the
method, such as delay-time and smoothing.

We can, however, go further and extract information which will be physically
useful.  To start, of great interest is the dimension $d$ of the inertial
manifold, \ie the manifold in which the dynamics evolves.  This dimension $d$ is obtained as follows.
From the fractal Lyapunov dimension $d_{_{L}}\approx$ 3.1 and the condition
that $d_{_{L}} < d$ (integer) we infer a {\sl lower} limit $4\leq d$.  On the
other hand, from our nonlinear analysis we have inferred an {\sl upper bound}
$d\leq d_e=4$.  {\sl We thus arrive at the important conclusion that the
dimension of the inertial manifold is indeed equal to 4}.

What do the 4 dimensions that we thus have found mean in simple terms?  A 4D
embedding implies, remarkably, that the behavior of the complicated, seemingly
erratic light-curve on any given day is {\sl uniquely} expressible in terms of
the behavior of the 4 preceding days only (taking $\Delta$=1 in eq.~1 for the
sake of argument).  In the context of a flow this is the number of independent
variables or ODEs which are sufficient to generate the observed signal.

The reader may then wonder why R~Sct is so different from, say a beat RR~Lyrae
star which, if it has 2 modes excited, would also be 4-dimensional.  The only
real difference is in the coefficients of the maps of the two types of stars:
the RR~Lyrae map produces an 2--torus (multi-periodic pulsation), whereas the
R~Sct map yields a strange attractor (chaotic trajectory).  The physical reason
why large amplitude chaotic behavior is not possible in RR ~Lyrae is that these
stars are only weakly nonadiabatic (the growth-rate to period ratio is of the
order of a percent), implying that amplitude modulations occur on a much longer
time-scale than the pulsation itself (mathematically, the dynamics evolves on a
near {\sl center manifold}, \cf below).  On the other hand, RV~Tau stars are
very far from adiabatic, a fact that allows modulations on the time-scale of
the period itself, obviously a necessary condition for chaos.

\vskip 10pt

In the literature {\sl first return maps} have been plotted for RV~Tau stars in
the hope that they might yield a clue as to the possible chaotic nature of
these objects (Veldhuizen \& Percy 1989; Saitou \etal 1989), but
disappointingly the first return maps turn out to be more or less scatter
diagrams.  In Figure~4 of Buchler, Koll\'ath and Serre (1995) we juxtapose the
first return map for our short R~Sct data set with a first return map of the
best synthetic signal, both constructed with the minima $\{m_i\}$ of the
light-curve (\ie a plot of $m_{i+1}$ \vs $m_i$).  The synthetic signal gives a
more compact plot, but which nevertheless has a great deal of structure, much
more than one obtains for example with the R\"ossler band (\eg Thompson \&
Stewart 1986).  This indicates that {\sl 1D return maps cannot capture} the
dynamics of R~Sct, and is a reflection of the differences in the Lyapunov
exponents and fractal dimensions of the two systems.

\titlea{PHYSICAL IMPLICATIONS}

We have demonstrated that the simplest and most natural explanation of the
pulsations of R~Sct is that they are produced by a 4-dimensional intrinsic
dynamics.  This dimension of four implies that 4 coordinates (or variables) are
sufficient, and necessary, to describe the dynamics.  The question is then how
this low dimension comes about, and what the physical nature of these four
variables is.

The hydrodynamics equations that describe the behavior of the star involve a
continuum (\eg $\rho(\vecr, t),$ $\vecu (\vecr,t),$ $s(\vecr,t),$ $\ldots$).
However, it is possible to project these equations onto the space of the normal
modes (linear eigenvectors) of which there is a denumerable infinity (a
similar, but much simpler example is afforded by a vibrating string or drum).
The {\sl configuration space}, in the parlance of WG94, is thus
infinite-dimensional for a pulsating star.

Our past theoretical work on Cepheids and RR~Lyrae stars (\eg Buchler 1993) has
shown that a description in terms of a few radial modes gives excellent
agreement with observations as well as with numerical hydrodynamical
simulations.  These particular stars are weakly 'nonadiabatic' (dissipative),
\ie the ratios of growth-rates to oscillation-frequencies are small for the
excited modes, at most of the order of a few percent, implying that the
dynamics evolves on a center manifold (erroneously called 'slow manifold' in
Buchler 1993).  Our understanding of this {\sl dimensional reduction}, from
infinite to a dimension of a few, is solidly based on center manifold theory
(\eg Guckenheimer \& Holmes 1983).

In contrast, W~Vir and RV~Tau stars are much more dissipative (large relative
growth-rates, of the order of tens of percent), and there is no longer a center
manifold.  However, both experiments and theory have shown that the dynamics of
dissipative fluids also often occur on low-dimensional manifolds (studied as
{\sl inertial manifold} by mathematicians, \eg Constantin \etal 1989).  Our
reconstruction analysis indicates that {\sl the dynamics of R~Sct evolves on
such an inertial manifold, of dimension} 4.

The existence of a 4D inertial manifold thus implies that all dynamical and
thermodynamical quantities throughout the star can be expressed in terms of
four basic variables.  The choice, or the nature of these 4 variables is not
unique of course (The variables are defined up to an arbitrary nonlinear
transformation).  However, in view of our above mentioned work on Cepheids and
RR~Lyrae stars it seems natural to identify these variables with modal
amplitudes, more specifically, with two vibrational (complex) amplitudes.
Strong direct support for this role of two vibrational modes comes from the
double spiral nature of the fixed points that we uncovered in the maps (\cf
\S4.5).  This then leads to the conclusion that {\sl the erratic pulsational
behavior is the result of the nonlinear interaction of two (complex)
mechanical modes of oscillation}.

In BSKM we have pointed out that the Fourier spectrum for the short data set
has noticeable power in the broad vicinity of 2.5 $\times$ the fundamental
frequency $f_0$ peak, even though a peak near $2.5 f_0$ may be less visible in
very long data sets (Koll\'ath 1990).  We have also recalled that
the period doubling cascade that led to chaos in our W~Vir model
pulsations originated in a parametric instability, caused precisely by a 5:2
resonance between the fundamental mode of oscillation of the stellar model and
an overtone (Moskalik \& Buchler 1990).  Whether for RV~Tau stars this
resonance also plays a role needs to be checked with a systematic numerical
hydrodynamic modelling survey of RV~Tau stars.

\titlea{\ \ CONCLUSIONS}

Chaos theory has had the scientific community excited for over a decade.
However, when applications to the real world are considered, it is often found
that very little useful information can be gained from such studies.  In a
recent article Ruelle (1994) has assessed the situation and he concludes that
the studies of the solar system have been perhaps the most fruitful application
of {\sl Hamiltonian chaos} theory.  It is therefore interesting that one of the
more fruitful applications of {\sl dissipative chaos} should also occur in
Astronomy, this time in variable stars.

The successful reconstruction of the dynamics of R~Sct shows that large
amplitude irregular stellar variability is no longer a mystery.  Our data
analysis corroborates the previous theoretical finding (BK87, KB88) that the
mechanism for the irregular variability of the Pop.  II Cepheids is
low-dimensional chaos that arises simply and naturally in the dynamics of the
star.

From a physicist's point of view it is interesting that the complicated and
relatively violent pulsational behavior of this star takes place in a 4-D
phase-space (inertial manifold).  We have produced evidence that the observed
chaotic behavior has a simple underlying physical model, namely it is the
result of the nonlinear interaction of two normal vibrational modes.

The type of analysis that we have presented here opens up the exciting
possibility of {\sl nonlinear astro-seismology}.  In a way very similar to the
study of multi-periodic stars (\eg bump or beat Cepheids, beat RR Lyrae,
pulsating white dwarfs, $\delta$ Scuti stars, \etc) the properties of the
irregular pulsations can be used to probe the stellar interior and to provide
new insights into stellar structure and evolution.  While it is clear that more
experience is required in the applications of the analysis one can foresee that
with a sufficient observing effort one may be able to extract systematically
{\sl quantitative information} from irregular variable star lightcurves.  This
could be in the form of quantities that are still relatively unfamiliar in
Astronomy, such as, for example, Lyapunov exponents and fractal dimensions.

The Pop. II Cepheids are as bright as their 10 day Pop. I counterparts and it
is conceivable that when we understand them better we may be able to use them
as well as standard candles for cosmological purposes.

\vskip 20pt

\centerline{\bfrm \ \ Acknowledgments}

\vskip 10pt

This research has been supported in part by NSF (AST92--18068 and
INT94--15868), a Hungarian OTKA grant (F4352), an RDA grant at UF, the French
Minist\`ere pour la Recherche et l'Espace, and RCI grant from IBM through UF.

\vfill\eject

\begref{References}

\ref
Abarbanel, H. D. I., Brown, R., Sidorowich, J. J., Tsimring, L. S. 1993,
Rev. Mod. Phys. 65, 1331 [ABST93]

\ref
Arp, H. C. 1955, AJ 60, 1.


\ref
Broomhead, D. S. \& King, G. P. 1987, Physica D 20, 217

\ref
Brown, R. 1992, Orthonormal Polynomials As Prediction Functions In
Arbitrary Phase-Space Dimensions, Institute for Nonlinear Science Preprint,
UC San Diego.

\ref 
Brown, R., Rulkov, N. F. \& Tracy, E. R. 1994. Phys. Rev. E 49, 3784.

\ref
Buchler, J. R. 1990, Ann. NY Acad. Sci. 617, 17

\ref
Buchler, J. R. 1993, Nonlinear Phenomena in Stellar
Variability, IAU Coll. 134, Eds. M. Takeuti \& J. R. Buchler 
(Dordrecht: Kluwer), repr. from 1993, Ap\&SS, 210, 1

\ref
Buchler, J. R., Koll\'ath, Z. \& Serre, T. 1995, in "Waves in
Astrophysics", Ann. NY Acad. Sci. (in press)

\ref
Buchler, J. R. \&  Kov\'acs, G. 1987, ApJLett. 320, L57-62 [BK87]

\ref
Buchler, J. R., Serre, T., Koll\'ath, Z. \& Mattei, J. 1995, Phys. Rev.
Lett. 74, 842 [BSKM].

\ref
Casdagli, M., Des Jardins, D., Eubank, S., Farmer J. D., Gibson, J., Hunter,
N. \& Theiler, J. 1992, in {\it Applied Chaos}, Ed. J. H. Kim \& J. Stinger,
335 ( N.Y.: Wiley)

\ref
Constantin, P., Foias, C. Nicolaenko, B. \& Temam, R. 1989. {\it Integral
Manifolds and Inertial Manifolds for Dissipative Partial Differential
Equations}, Appl. Math. Sci. 70 (N.Y.: Springer)

\ref
Gingold, R. A. 1974, ApJ 193, 177

\ref
Glendenning, P. \& Tresser, C. 1985, J. Physique Lett. 46, L347

\ref
Gouesbet, G. 1991, PR 43A, 5321.

\ref Guckenheimer, J. \& Holmes, P. 1983, {Nonlinear Oscillations, Dynamical
Systems and Bifurcation Theory}, (N.Y.: Springer)

\ref
Koll\'ath, Z. 1990, MNRAS 247, 377.

\ref
Koll\'ath, Z., Buchler J. R., Serre, T. \& Mattei, J. 1995, ApJ 
      (in preparation).

\ref
Koll\'ath, Z. \& Szeidl, B. 1993, A\&A 277, 62.

\ref
Kov\'acs, G. \& Buchler, J. R. 1988, ApJ, 334, 971, [KB88]. 

\ref
Kukarkin, B. V.  1975, {\it Pulsating Stars} (N.Y.: Wiley)

\ref
Ludendorff, H. 1928, {\it Handbuch der Astrophysik}, 6, 49

\ref
Moskalik, P. \& Buchler, J. R. 1990, ApJ 355, 590.

\ref
Ott, E. 1993, {\it Chaos in Dynamical Systems} (Cambridge: Univ. Press), [O93]

\ref
Press, W. H., Teukolski, S. A., Vetterling, W. T. \& Flanney, B. P. 1992,
{\it Numerical Recipes} (University Press: Cambridge).

\ref
Reinsch, C. H. 1967, Numerische Mathematik, 10, 177

\ref
Ruelle, D. 1994, Physics Today, July 24

\ref 
Saitou, M., Takeuti, M. \& Tanaka Y. 1989, PASJ 41, 297


\ref
Scargle, J. D. 1981, ApJS 45, 1

\ref
Serre, T., Koll\'ath, Z. \& Buchler, J. R. 1995a, A\&A (submitted) [SKB]

\ref
Serre, T., Koll\'ath, Z. \& Buchler, J. R. 1995b, A\&A (submitted)

\ref
Thompson, J. M. T. \& Stewart, H. B. 1986, {\it Nonlinear Dynamics and Chaos}
(John Wiley ans Sons)

\ref
Veldhuizen, T. \& Percy,  J. R. 1989. {\it J. AAVSO} 18, 97.

\ref
Weigend, A. S. \& Gershenfeld, N. A. 1994, {\it Time Series Prediction} 
 (Reading: Addison-Wesley) [WG94]

\end
\bye